\documentclass{elsarticle}
\usepackage{natbib}
\usepackage{amsmath, amssymb, amsfonts}
\usepackage{latexsym}
\usepackage{epsfig}
\usepackage{lineno,hyperref}
\usepackage{color}
\modulolinenumbers[5]
\journal{Journal of \LaTeX\ Templates}
\def\blue#1{{\color{blue}#1}}
\definecolor{dag}{rgb}{0.1,0.4,0.1}

\newcommand{\tr}{^{\prime}}
\newcommand{\diag}{{\rm diag}}
\def\m#1{\mbox{#1}}                
\def\bd#1{\mbox{\boldmath $#1$}}    
\def\cg#1{\mbox{${\cal #1}$}}
\def\cgl#1{\mbox{\scriptsize {${\cal #1}$}}}
\newtheorem{theorem}{Theorem}
\newtheorem{lemma}{Lemma}

\newtheorem{proposition}{Proposition}

\newtheorem{remark}{Remark}
\newtheorem{corollary}{Corollary}

\biboptions{authoryear}

\begin{document}

\begin{frontmatter}

\title{An extended class of RC association models: \\
estimation and main properties}

\author{Antonio Forcina, Maria Kateri}
\address{Dipartimento di Economia, University of Perugia, Italy; Institute of Statistics, RWTH Aachen University, Germany}
\ead{forcinarosara@gmail.com, maria.kateri@rwth-aachen.de}
\begin{abstract}
The extended class of multiplicative row-column (RC) association models, introduced in this paper for two-way contingency tables, allows users to select both the type of logit (local, global, continuation, reverse continuation) suitable for the row and column classification variables and the scale on which interactions are measured. As in \cite{Kateri95} for the case  of local logits, our extended class of bivariate interactions is linked to divergence measures  and, by means of a representation theorem, we provide reconstruction formulas for the joint probabilities depending on pairs of logit types.
These results are the key to show that, given marginal logits, our extended interactions determine uniquely the bivariate distribution. We also determine the kind of positive association which is implied by our extended interactions being non negative. Quick model selection within this wide class can be performed by an efficient algorithm for computing maximum likelihood estimates which exploits the properties of a reduced rank constraint imposed on the matrix of extended interactions and allows for additional linear constraint on marginal logits. An application to social mobility data is presented and discussed.

\end{abstract}
\end{frontmatter}
%
\section{Introduction}
The RC association models introduced by \cite{goodman1979simple} as a flexible class of models for
investigating association in a two-way contingency table, were, later, extended by \citep{Goodman81}
and \cite{gilula1988ordinal} to include models of Correspondence Analysis which had been developed,
initially, with partially different objectives. Later, \cite{Kateri95} formulated a general class of RC association models that includes the original association models and the correspondence analysis model
as special cases. They also showed that each model in their class is optimal with respect to a
divergence measure which determines the form of a class of generalized interactions.

An extension in a different direction was proposed by \cite{Bart02} who noted
that the original RC association models could be seen as defined by a rank
constraint on the matrix of local log-odds rations; they suggested that a
similar rank constraint could be imposed to the matrix of log-odds ratios (LORs
for short in the following) defined by choosing logits
of a general type (local, global, continuation, reverse continuation) for the
row and column variable. The motivation for this extension is that logits of
a certain type may be more appropriate than the LORs based on adjacent row
and columns categories of a two way table. These choices may be combined as
in \cite{douglas1990positive} leading to a variety of interaction parameters
which can be subjected to a rank constraint. Thus, for each chosen pair of
logits, a corresponding RC association model may be defined and fitted by
imposing, if suitable, additional linear constraints on marginal logits and
LORs. More recently, \citet[Chapter 5]{EspeM} has proposed an extension of
RC association models which combines certain generalized LORs of different
types with the function used by \cite{Kateri95} to define $\phi$-scaled
interactions.

This paper defines a new class of interaction parameters for two-way tables obtained by combining the
work of \cite{Kateri95} and \cite{Bart02} and may be seen as a generalization of \cite{EspeM}.
In particular, selecting a specific divergence measure, the type of logit for the row and column variable and the rank of the corresponding matrix of interactions, an extended RC association model may be defined. An algorithm for maximum likelihood estimation (ML) which implements the results on rank constraints described by \cite{Bart02} is presented.
An adaptation of the {\it regression}
version of the \cite{aitchison1958maximum} approach to constrained ML estimation described by
\cite{evans2013two} is implemented together with a suitable line search for improving stability.

The flexibility and the potentials of the new class of models is illustrated
in the application where, for convenience, we restrict to \cite{cressie1984multinomial} power divergence measures. This is a parametric family of divergences and enables thus the definition of a large number of alternative models by simply controlling the value of the power parameter in the divergence measure.

Concerning the theoretical properties of the new models, by extending the results in \cite{Kateri95}, we show that they may be derived by solving a constrained optimization problem.
This result is the key to show that, given row and column margins, the new interaction parameters determine uniquely the bivariate joint distribution. For the case when both classification variables are ordinal, for each pair of logit types, we determine the form of positive association as in \cite{douglas1990positive}\blue{,} which is implied by all extended interactions being non negative

The paper is organized as follows. The working framework and some
preliminaries on generalized logits, LORs and association models are
presented in Section 2. The new extended association models are defined in
Section 3 and some results on optimality properties and positive association
are presented in Section 4. An algorithm for ML estimation is
presented in Section 5 and an application to social mobility tables is
discussed in Section 6.
\section{Preliminaries}
Consider an $I_1\times I_2$ contingency table containing the observed
frequencies for $X_1,X_2$, two categorical variables, corresponding to the
rows and columns of the table, respectively. Let $\pi_{ij}$ denote the $i$-th
row and $j$-th column cell probability, for $i\in \cg I=\{1,\ldots,I_1\}$ and
$j\in \cg J=\{1,\ldots,I_2\}$, while $\bd\pi$ denotes the vector of cell
probabilities, with the columns categories running faster.  Furthermore let
$\bd y$ denote the corresponding vector of observed frequencies. Finally, let
$\bd\pi_1=(\pi_{1+}, \ldots, \pi_{I_1+})$ and $\bd\pi_2=(\pi_{+1}, \ldots,
\pi_{+I_2})$ be the vectors of marginal row and column probabilities.

\subsection{Generalized logits and log odds ratios}\label{generalized_LORs}
There are four main types of logit  considered in the literature: local (L),
global (G), continuation (C) and reverse continuation (R). Denote with
$l_{\ell}$, $\ell=1,2$, the logit type considered appropriate for
$X_{\ell},\: \ell=1,2$, respectively where $l_{\ell} \in \{\m{L, G, C, R}\}$;
in the following we assume that the type of logit used to define marginal
parameters for a given variable is also used when computing conditional
logits and log-odds ratios (LORs for short). To define logits and LORs, it is
convenient to introduce a class of events and marginal distributions over
these events. For $\ell=1,2$, let $E(x_{\ell},b_{\ell},l_{\ell})$ denote a
class of events determined by the cut point $x_{\ell}<I_{\ell}$, the binary
indicator $b_{\ell}$ and the logit type $l_{\ell}$ according to the following
rules
\begin{align*}
E(x_{\ell},0,l_{\ell}) &= \left\{
\begin{array}{l} x_{\ell} \m{ if $l_{\ell}$ is L or C}\\
1,\dots x_{\ell} \m{  if $l_{\ell}$ is G or R}
\end{array}\right. \\
E(x_{\ell},1,l_i) &= \left\{
\begin{array}{l} x_{{\ell}+1} \m{ if $l_{\ell}$ is L or R} \\
x_{{\ell}+1},\dots , I_{\ell} \m{  if $l_{\ell}$ is G or C}
\end{array}\right.
\end{align*}
Furthermore let $\cg M$ be the set of indices defining a marginal
distribution and
$$
p_{\cgl M}(\bd x_{\cgl M}; \bd b_{\cgl M};\bd l_{\cgl M})
= P(X_\ell\in E(x_\ell,b_\ell,l_\ell), \forall \ell\in \cg M).
$$
In the given context, $\cg M$ can be 1 for the row marginal, 2 for the column
marginal and (1,2) for the joint distribution; the $(1,2)$ suffix however can
be omitted without causing ambiguities. For convenience, we recall the
definition of the different logits applied to the marginal distribution of
$X_1$:
\begin{center}
\begin{tabular}{clclcl}
    L: & $\eta_{1;L}(i)$ & = &  $\log p_1(i;1;L)-\log p_1(i;0;L)$ &=& $\log
      \pi_{i+1,+}-\log\pi_{i+}$, \\
    G: & $\eta_{1;G}(i)$ & = & $\log p_1(i;1;G) - \log p_1(i;0;G)$ &=& $
     \log\sum_{i+1}^{I_1} \pi_{h+}- \log \sum_1^i\pi_{h+}$, \\
    C: & $\eta_{1;C}(i)$ & = & $\log p_1(i;1;C) - \log p_1(i;0;C)$ &=&
      $\log\sum_{i+1}^{I_1} \pi_{h+}- \log \pi_{i+}$,\\
    R: & $\eta_{1;R}(i)$ & = & $\log p_1(i;1;R) - \log p_1(i;0;R)$ &=&
      $\log\pi_{i+1,+}- \log \sum_1^i\pi_{h+}$,
\end{tabular}
\end{center}

As shown in \citet[Proposition 1]{bartolucci2007extended}, LORs for a given
pair of logit types may be computed directly as second order contrasts of log
probabilities or as difference between logits of either variable conditional
on the other belonging to the disjoint subsets $E(x_{\ell},1,l_{\ell})$ and
$E(x_{\ell},0,l_{\ell})$. For instance, the expressions for LORs of type CG
may be computed as
\begin{align*}
\eta_{1,2;C,G}(i,j) = & [\log
p(i,j;1,1;C,G)-\log p(i,j;0,1;C,G)]-\\ & [\log p(i,j;1,0;C,G)-\log
p(i,j;0,0;C,G)].
\end{align*}
Among the 16 possible combinations of logit types (sf. \cite{douglas1990positive}),
the most known LORs are the {\it local} (LL), {\it cumulative} (LG), {\it global} (GG) and
{\it continuation} (LC).
It is well known
that for an $I_1\times I_2$ contingency table the set of $(I_1-1)(I_2-1)$
LORs for  a given pair of logit types captures the underlying association
structure. Furthermore, this minimal set together with the row and column
marginal distributions define the table uniquely, a result that follows, for
instance, from Theorem 1 in \cite{bartolucci2007extended}. Thus, a model can
be defined either in terms of the cell probabilities or in terms of the
minimal set of LORs and the marginal logits. Hence, it is
obvious that the role of the LORs is fundamental in modeling
contingency tables. It is worth recalling that in classical log-linear models
the interaction parameters are defined in terms of the LORs of type L-L.
\subsection{Defining association models by a rank constraint}
The original formulation by Goodman, followed in most treatments of association models, assumes that
\begin{equation}
\log \pi_{ij} = \zeta+\alpha_i+\beta_j+\sum_1^K \psi_k\mu_{k,i}\nu_{k,j}, \ \
 i\in \cg I, \ j\in \cg J,
\label{Goodman}
\end{equation}
where $K\leq\: K^*= min(I_1, I_2)-1$ and, for each $k$, $\psi_k$, $\bd\mu_k=(\mu_{k,1},\ldots,\mu_{k,I_1})$
and $\bd\nu_k=(\nu_{k,1},\ldots,\nu_{k,I_2})$ are, respectively, the association parameter, and the vectors of row and column scores; in addition,  the $\alpha$s, $\beta$s, $\mu$s and $\nu$s are subject to suitable identifiability constraints. Model (\ref{Goodman}) is denoted as RC($K$), with RC($K^*$) being the saturated model.

The association models can equivalently be defined in terms of certain restrictions imposed on the $(I_1-1)\times(I_2-1)$ matrix of local LORs $\bd M(LL)=(\eta_{1,2;L,L})$. It can be shown \citep[see][Chapter 6]{Kateri14}, that, for given marginals $\bd\pi_1$ and $\bd\pi_2$  model (\ref{Goodman}) is equivalent to
\begin{equation}\label{RC_LOR}
\eta_{1,2;L,L}=\log\left(\frac{\pi_{ij}\pi_{i+1 j+1}}{\pi_{i,j+1}\pi_{i+1 j}}\right) =
\sum_k\psi_k(\mu_{k,i+1}-\mu_{k,i})(\nu_{k,j+1}-\nu_{k,j}),
\end{equation}
for all $i< I_1, j< I_2$.

The definition of association models in terms of cell probabilities (\ref{Goodman}) is more familiar,
 in the log-linear setup. On the other hand, interpretation of the results is based on odds ratios and thus expression (\ref{RC_LOR}) is preferable from this point of view.
Furthermore, for the estimation of (\ref{Goodman}), as discussed next, only the rank $K$ of matrix $\bd M(LL)$ is required.
The parameters $\psi_k, \:\mu_{k,i},\: \nu_{k,j}$ can be easily computed once the model has been estimated via a singular value decomposition (SVD) approach. Expression (\ref{RC_LOR}) is also convenient for extending association models, since by imposing the same restrictions to different types of LORs, alternative association models arise, as discussed in \cite{Bart02}; see also \citet[Chapter 7]{Kateri14}.

The estimation procedure for RC($K$) corresponding to a pair of logits $l_1 l_2$, is based on the well-known fact that the marginal distributions of the row and column variables are determined by $I_1-1$ and $I_2-1$ marginal logits respectively, while the association depends on the corresponding matrix of interactions $\bd M(l_1l_2)=(\eta_{1,2;l_1,l_2})$. It can be easily shown that an RC($K$) model defined on a contingency table with fixed the associated marginal logits, is equivalent to assuming that the corresponding matrix $\bd M(l_1l_2)$ has rank $K$.
\cite{Bart02} proposed an estimation procedure for RC($K$) models that imposes directly the appropriate rank constraint on
$\bd M(l_1l_2)$. A method for implementing the rank constraint is presented in Section 5.
\subsection{The $\phi$-divergence and the Cressie and Read power divergence}
In a statistical information theoretic set-up, \cite{Kateri95} defined a generalized class of dependence models for two-way contingency tables, extending (\ref{Goodman}) through the family of $\phi$-divergence measures \citep[see][]{csiszar1967information}. If $\bd\pi$ denotes a discrete finite bivariate probability distribution, in vector form, and $\bd\pi_0$ is the vector of joint probability for a reference distribution, then the $\phi$-divergence measure between $\bd\pi$ and $\bd\pi_0$ is defined as
\begin{equation}
I_{\phi}(\bd\pi,\bd\pi_0) = \sum_v \pi_{0v} \phi\left(\frac{\pi_v}{\pi_{0v}}\right) ,
\label{phi-div}
\end{equation}
where $\phi$ is a real--valued convex function on $[0,\infty)$ with $\phi(1)=\phi'(1)=0$, $0\phi(0/0)=0$, $0\phi(y/0)=\lim_{x\rightarrow\infty} \phi(x)/x$. Setting $\phi(x)=x\log x-x+1$, (\ref{phi-div}) is reduced to the Kullback--Leibler (KL) divergence while for $\phi(x)=\frac{x^{\lambda+1}-x-\lambda(x-1)}{\lambda(\lambda+1)}$ it becomes the power divergence measure of \cite{cressie1984multinomial}
\begin{equation}
I_{CR}(\bd\pi,\bd\pi_0;\lambda) =
\sum_v \pi_{0v}\frac{1}{\lambda(\lambda+1)} \left[
\left(\frac{\pi_v}{\pi_{0v}}\right)^{\lambda+1}-\frac{\pi_v}{\pi_{0v}}\right],  \ \ \lambda \ne 0, -1,
\label{CRead}
\end{equation}
which is a very important parametric family of divergence measures and on which we will focus later on.
When $\lambda$ tends to 0, it converges to the KL divergence measure. For more technical features of these measures see for instance \cite{Pardo06} or \cite{Cressie2000}.
In the following it will be convenient to denote the first derivative of function $\phi$ by $F$ and the inverse of $F$ by $G$, i.e. $F=\phi'$ and $G=F^{-1}$ which, as shown by \cite{Kateri95} (proof of Theorem 2.1), exists everywhere.
\section{A new class of extended RC association models}
The class of models defined in this section tries to combine logits of a given type, not necessarily the same on the row and column marginal distributions, with a new class of interaction parameters which, once arranged into a $(I_1-1)\times(I_2-1)$ matrix satisfy a rank constraint as in an ordinary RC model.
We first define the new interaction parameters and then show that, together with the set of marginal logits, they determine uniquely a bivariate distribution.
\subsection{A $\phi$-divergence scaled class of interaction parameters}
The $\phi$-divergence generalized association model of \cite{Kateri95}, which refer to ORs of the type L-L, are based on the $F$ scaled ratios $F\left(\frac{\pi_{ij}}{\pi_{i+}\pi_{+j}}\right)$ for each cell, i.e. their $F$-scaled deviance from independence instead their $\log$-scaled, which is the scale of the standard local LORs. Note that the log-scale corresponds to the $F$-scale for the KL divergence measure \citep[see also][]{Kateri18}. In order to consider $\phi$-divergence generalized association models based on other types of ORs, we need to extend the ratios $\frac{\pi_{ij}}{\pi_{i+}\pi_{+j}}$ to
$$
\varrho_{ij}(u,v;l_1,l_2)=\frac{p(i,j; u,v; l_1,l_2)}{p_{1}(i; u; l_1) \:p_{2}(j; v);
l_2)},
$$
where $i\in\cg I$, $j\in\cg J$ and $u,v$ are binary indicators.

Then an extended class of interaction parameters for two-way contingency tables may be defined as
\begin{eqnarray}\label{gamma_s}
\gamma_{ij}(F;l_1,l_2) =&  F[\varrho_{ij}(1,1;l_1,l_2)] - F[\varrho_{ij}(1,0;l_1,l_2)] \\
 -& F[\varrho_{ij}(0,1;l_1,l_2)] + F[\varrho_{ij}(0,0;l_1,l_2)]. \nonumber
\end{eqnarray}
where the function $F(u)$, for the power divergence measure, may be simplified to $u^\lambda/\lambda$
when $\lambda\ne 0$.  Imposing on the $F$-scaled interaction parameters (\ref{gamma_s}) the structure of an RC model, i.e. a suitable rank constraint, an extended RC(K) model can be defined, which is expressed in terms of the $\gamma_{ij}$s in (\ref{gamma_s}) as
\begin{equation}
\gamma_{ij}(F;l_1,l_2) = \sum_k\psi_k (\mu_{k,i+1}-\mu_{k,i})
(\nu_{k,j+1}-\nu_{k,j}), \ \ i<I_1, j<I_2,
\label{GeKat_gamma}
\end{equation}
which is an extension of (\ref{RC_LOR}).
\begin{remark}
Even in the context of $F$-scaled interactions, RC(1) models, in addition to being the most useful in practice, are also those whose parameters have a more clear interpretation. However, in the following, theoretical results will be stated for the general RC(K) model.
\end{remark}
Under (\ref{GeKat_gamma}) the $\phi$-generalized interaction parameters $\gamma_{ij}$s have exactly the same form as the corresponding LORs under the standard association models.
For certain pairs of logit types, the interaction parameters (\ref{gamma_s}) have been introduced in
Definition 5.1.1 of \cite{EspeM} while special types of model expression (\ref{GeKat_gamma}) are discussed in his Chapter 6.
\subsection{A representation theorem}\label{represent_section}
A representation theorem for the subclass of models where both logit types are set to L was derived  by \cite{Kateri95}. It may be convenient to summarize their results before presenting an extension to the more general class of models introduced in this paper. They showed that, within the class of all bivariate distributions $\bd \pi$ having the same row and column marginals, the same set of row and column scores $\bd \mu,\:\bd \nu$ and the same value of the correlation coefficient $\rho$ = $(\bd\mu\otimes\bd\nu)\tr \bd\pi$, the distribution which minimizes the discrepancy in (\ref{phi-div}) relative to independence must have the form
\begin{equation}
\pi_{ij} = \pi_{i+} \pi_{+j} G(\alpha_i + \beta_j +\psi \mu_i\nu_j), \ \
i\in \cg I, \ j\in \cg J.
\label{KaThe}
\end{equation}
This result may be interpreted as saying that, if we believe that the appropriate divergence measure should be the one in (\ref{phi-div}), or the one in (\ref{CRead}) for a given $\lambda$, then an appropriate $\phi$-generalized RC association model should have the form prescribed by (\ref{KaThe}).

Though the above result is stated for an RC(1) model, it can be easily extended to an RC($K$), for $1\leq K\leq K^*$, as discussed in the final section of \cite{Kateri95}.
An extension of (\ref{KaThe}) to the class of RC models introduced in this paper is presented below; however, because, as we shall see, the case of logits of type  L requires a different treatment, we first consider the case where the logits for rows and columns are both different from L.

Let $r(i,j;u,v)$ = $p_{1}(i; u) p_{2}(j; v)$, let $\phi$ be the function that characterize a certain divergence measure and define
$$
I_{\psi} = \sum_{ij}\sum_{uv} r(i,j;u,v) \phi\left(\frac{p(i,j;u,v)} {r(i,j;u,v)}\right),
$$
this may be interpreted as a discrepancy measure computed within each $2\times 2$ probability table
determined by logit types and a pair $i,j$ of cut points and then summed across all possible pairs of cut points.
Suppose also that we have $K$ sets of row and column scores $\mu_{k,i},\:\nu_{k,j}$, $k=1,\dots ,K$ and define
$$
\rho_k = \sum_{ij}\sum_{uv} p(i,j;u,v)\mu_{k,i+u}\nu_{k,j+v};
$$
for given row and column scores, this may be interpreted as a measure of covariance within each $2\times 2$ table determined again by each pair of cut points and summed across all possible pairs of cut points. Let also
\begin{equation}\label{q_marginal}
q(i,j;u,+)= p(i,j;u,0)+p(i,j;u,1), \: q(i,j;+,v)= p(i,j;0,v)+p(i,j;1,v);
\end{equation}
these are the sums by row and column within each $2\times 2$ table determined by cut points and logit types.
\begin{theorem}
For an $I_1\times I_2$ probability table suppose that both the logit types adopted are different from L; within the class of all possible bivariate distributions having the same set of marginals $q(i,j;u,+)$ and $q(i,j;+,v)$ relative to $p(i,j; u,v)$, the same set of row and column scores $\mu_{k,i+u},\: \nu_{k,j+v}$ and the values of $\rho_k$ fixed at $\rho_k^*$,  for $i\in\{1,\ldots,I_1-1\}$, $j\in\{1,\ldots,I_2-1\}$, $u,v\in\{0,1\}$,
the distribution closest to independence in terms of $I_{\psi}$ has the form
\begin{equation}
p(i,j; u,v) = p_{1}(i; u) p_{2}(j; v) G\left[\alpha_{iju} +\beta_{ijv}+
\sum_k\psi_k \mu_{k,i+u}\nu_{k,j+v}\right],
\label{GeKat}
\end{equation}
where the function $G$ is the inverse of the function $F$.
\label{ProKat}
\end{theorem}
{\sc Proof}. Define the lagrangian
\begin{align*}
L =& I_{\psi}-\sum_{iju}\alpha_{iju} \left[ p(i,j;u,0)+p(i,j;u,1)-q(i,j;u,+)\right] \\
-& \sum_{ijv}\beta_{ijv} \left[p(i,j;0,v)+p(i,j;1,v)-q(i,j;+,v)\right] -
\sum_k\psi_k(\rho_k-\rho_k^*)
\end{align*}
and differentiate with respect to $p(i,j;u,v)$ for $i<I_1-1$, $j<I_2-1$, $u,\: v$,
$$
\frac{\partial L}{\partial p(i,j;u,v)} = F\left(\frac{p(i,j;u,v)} {q_{ij}(u,v)}\right)
-\alpha_{iju}-\beta_{ijv}-\sum_k\psi_k \mu_{k,i+u}\nu_{k,j+v}\ =0,
$$
which implies (\ref{GeKat}) because an inverse of $F$ is assumed to exist everywhere. $\Box$

Now suppose that the logit type for rows is set to L while that for columns is any among G, C, R and note that the quantities $p(i,j,1,v)$ and $p(i+1,j,0,v)$ are equal for all $i=1,\dots , I_1-2$. To take this into account, we consider as variables in the optimization problem the probabilities $p(i,j,v)$ which are set to $p(i,j,0,v)$ if $i < I_1$ and to $p(i-1,j,1,v)$ when $i=I_1$; this is equivalent to consider a collection of $I_1\times 2$ tables, one for each $j<I_2$. The discrepancy to be minimized may be written as
$$
\sum_{j=1}^{I_2-1}\left[\sum_{i=1}^{I_1} \sum_v r(i,j,v)\phi(p(i,j,v)/r(i,j,v)) \right],
$$
where $r(i,j,v)$ = $\pi_{i+}\sum_i p(i,j,v)$. Redefine also the covariances as a sum of the covariances for fixed $j$ within each $I_1\times 2$ tables and summed over $j$
$$
\rho_k=\sum_{j=1}^{I_2-1}\sum_i\sum_v \mu_{k,i}\nu_{k,j+v} p(i,j,v)
$$
For the marginal constraints, suppose we know the row margins $\pi_{i+}$ and the column totals for fixed $j$, $p(+,j,v)$ = $\sum_{i=1}^{I_1} p(i,j,v)$. Thus we must have
\begin{align*}
\sum_v p(i,j,v) &= \pi_{i+}, \quad i=1,\ldots, I_1, \quad j=1,\ldots, I_2-1,\\
\sum_i p(i,j,v) &= p(+,j,v), \quad j=1,\ldots, I_2-1, \quad v=0,1.
\end{align*}
The first set of constraints depends on the pair $i,j$, so the Lagrange multipliers might be called $\alpha_{ij}$ while the second set of constraints depend on the pair $j,v$, so call it $\beta_{jv}$ the corresponding Lagrange multipliers. Thus,
analogously to Theorem \ref{ProKat}, we may state the following
\begin{corollary}
For $I_1\times I_2$ probability tables where the logit type for rows is L and that for columns is different from L, the distribution closest to independence which has a prescribed set of covariances $\rho_k$, $k=1,\dots ,K$ and the marginals as specified above must have the form
\begin{equation}\label{GeKat2}
p(i,j,v) = \pi_{i+} p(+,j,v) G\left(\alpha_{ij}+\beta_{jv}+ \sum_k\psi_k\mu_{k,i}\nu_{k,j+v}\right),
\end{equation}
\label{LG}
where $i\in\cg I$, $j\in\{1,\ldots,I_2-1\}$ and $v\in\{0,1\}$.
\end{corollary}
\section{Properties of the new class of models}
\subsection{Uniqueness of the Mapping}
We have seen in Section \ref{generalized_LORs} that an important property of LORs (and consequently of association models which impose a rank constraint on LORs) is that the minimal set of LORs, along with the row and column marginal distributions, define a probability table uniquely. It is thus important to verify whether, for the new introduced extended RC models, an analogue mapping between appropriately defined bivariate distributions and the corresponding minimal set of $F$-scaled LORs and corresponding row and column totals is one to one.
Because the proof that such a mapping is one to one is based on the form of the representation theorem, it is convenient to consider the three cases discussed in Section \ref{represent_section} separately. In addition, as we will argue below, logits of type C and R need additional care.
Below, to simplify notations, we replace $\gamma_{ij}(F;l_1,l_2)$ with $\gamma_{ij}$ because the $F$ function and the logit types, once chosen, are kept fixed; further we set
$\upsilon_{ij}$ = $\sum_k\phi_k\mu_{k,i}\nu_{k,j+v}$.
\subsubsection{Logits of type LL}
The following is a rephrased version of Theorem 5.3.1 in \cite{EspeM}; the proof we provide is shorter but uses, essentially, the same ideas.
\begin{proposition}
Let $\bd\pi$ be the vector of cell probabilities of an $I_1\times I_2$ contingency table. For fixed row and column marginal distributions, $\bd\pi_1$ and $\bd\pi_2$, respectively, the vector $\bd\gamma$ of $\phi$-scaled interactions of type LL determines uniquely the joint distribution $\bd\pi$.
\end{proposition}
{\sc Proof}
Recall that, from the representation Theorem,
\begin{equation}
\pi_{ij} = \pi_{i+}\pi_{+j}G(\alpha_i+\beta_j+\upsilon_{ij}), \ i\in \cg I, \ j\in \cg J,
\label{base}
\end{equation}
while by definition $\bd\gamma$ = $\bd C\bd\upsilon$, where $\bd C$ = $\bd C(I_1)\otimes \bd C(I_2)$ and $\bd C(h)$ is the $(h-1)\times h$ adjacent difference operator. Uniqueness would not hold if it existed $\tilde{\bd\pi} \ne\bd\pi$, having the same marginal distributions as $\bd\pi$, $\bd\pi_1$ and $\bd\pi_2$, but depending on $\tilde{\bd\upsilon} \ne\bd\upsilon$ such that
$\tilde{\bd\gamma}=\bd\gamma$, i.e. $\bd C(\bd\upsilon-\tilde{\bd\upsilon})$ = $\bd 0$.
Consider the $I_1\times I_2$ table $\bd\Delta$ with entries
$$
\Delta_{ij}=\frac{\pi_{ij} - \tilde\pi_{ij}}{ \pi_{i+}\pi_{+j}}  = G(\alpha_i+\beta_j+\upsilon_{ij})-
G(\tilde\alpha_i+\tilde\beta_j+\tilde\upsilon_{ij});
$$
then we may write $\pi_{ij} - \tilde\pi_{ij}$ = $\pi_{i+}\pi_{+j} \Delta_{i,j}$.
The structure of the $\bd C$ matrix implies that $\upsilon_{ij}-\tilde\upsilon_{ij}$ = $a_i+b_j$ for some $a_i,b_j$,
leading to
$$
\Delta_{ij} = G(\alpha_i+\beta_j+\upsilon_{ij})-
G(\alpha_i^*+\beta_j^*+\upsilon_{ij}),
$$
where $\alpha_i^*=\tilde\alpha_i-a_i$ and $\beta_j^*=\tilde\beta_j-b_j$.
Because of the marginal constraints, the $\Delta_{ij}$ have to satisfy the following set of equality constraints
\begin{equation}
a)\: \sum_i\pi_{i+} \Delta_{ij} = 0,\forall j\in\cg J,\quad
b) \: \sum_j\pi_{+j} \Delta_{ij} = 0, \forall i\in\cg I. \label{RCm}
\end{equation}
Since the row and column marginal probabilities are all positive, the constraints in (\ref{RCm}) imply the following rule of sign change (sc) for the entries of table $\bd\Delta$:
\begin{description}
\item[(sc)] within each row $i$ (and column $j$), unless uniqueness holds (i.e. all $\Delta_{ij}$ = 0),  there must be, at least, a negative and a positive element.
\end{description}
Suppose that, for some $i,j$, $\Delta_{ij}$ is, say, negative, then
$$
G(\alpha_i+\beta_j+\upsilon_{ij}) < G(\alpha_i^*+\beta_j^*+\upsilon_{ij}).
$$
Since the function $G$ is strictly increasing, applying its inverse to both sides of the equation above, it can be easily seen that $\Delta_{ij}$, if different from 0, has the same sign as
$$
D_{ij} = \alpha_{i}-\alpha_{i}^*+ \beta_j-\beta_j^*.
$$
We show next that, unless all $D_{ij}$ = 0, their additive structure is incompatible with the (sc) rule. Let $h,\:k,\:r, \:s$, be indices such that
\begin{align*}
\alpha_{h}-\alpha_{h}^*\leq & \alpha_{i}-\alpha_{i}^*\leq
\alpha_{k}-\alpha_{k}^*,\forall i,\\
\beta_r-\beta_r^*\leq &\beta_j-\beta_j^* \leq
\beta_s-\beta_s^*, \forall j.
\end{align*}
Unless uniqueness holds (all $D_{ij}=0$), $D_{hr}$ must be negative and (sc) for row $h$ implies that
$D_{hs}=\max_{j \in \cg J}\{D_{hj}\}$ must be positive. But $D_{hs}$ is the smallest value in column $s$, so all other terms must also be positive, which violates (sc) in column $s$;  thus uniqueness must hold. $\Box$
\subsubsection{Row and column logits different than L}
\begin{proposition}
Let $\bd p(i,j)$,$i\in\{1,\ldots, I_1-1\}$ and $j\in\{1,\ldots, I_2-1\}$, be the set of $2\times 2$ tables with entries as defined in (9) and both logit types different than L. For fixed corresponding row and column marginals, $\bd p_1(i)=(p_1(i;0), p_1(i;1))$ and $\bd p_2(j)=(p_2(j;0), p_2(j;1))$, the interactions
$$\gamma_{ij} = \upsilon_{ij}(0,0)-\upsilon_{ij}(0,1)- \upsilon_{ij}(1,0)-\upsilon_{ij}(1,1) , $$
within each $2\times 2$ table, where $\upsilon_{ij}(u,v) = \sum_k\varphi_k \mu_{k,i+u}\nu_{k,j+v}$, $u,v \in\{0,1\}$,
determine uniquely the tables $\bd p(i,j)$.
\end{proposition}
{\sc Proof}
Let us start with the GG case.
Within each $2\times 2$ table for fixed $i,j$, the situation is analogous to that in the LL case in the sense that both the difference $\upsilon_{ij}(u,v)-\tilde\upsilon_{ij}(u,v)$ and the pair $\alpha_{iju},\:\beta_{ijv}$ are additive with respect to $u,v$. In analogy with the LL case, we may define
\begin{align*}
\Delta_{ij}(u,v)= &\frac{p(i,j;u,v) - \tilde p(i,j;u,v)}{ p_1(i;u)p_2(j:v)}  \\
=& G(\alpha_{iju}+\beta_{ijv}+\upsilon_{ij}(u,v))-
G(\alpha_{iju}^*+\beta_{ijv}^*+\upsilon_{ij}(u,v))
\end{align*}
and $D_{ij}(u,v)$ = $\alpha_{iju}-\alpha_{iju}^*+\beta_{ijv}-\beta_{ijv}^*$. In this case the marginal constraints may be written as
$$
a) \: \sum_u p_1(i;u)\Delta_{ij}(u,v)=0,\quad
b) \: \sum_v p_2(j;v)\Delta_{ij}(u,v)=0,
$$
which hold because, for instance, $p(i,j;u,+)$ = $\tilde p(i,j;u,+)$ = $p_1(i;u)$. Arguing as in the LL case based on the strict monotonicity of function $G$, $\Delta_{ij}(u,v)$ and $D_{ij}(u,v)$, if different from 0, must have the same sign. Furthermore, it can be shown that the marginal constraints within each $2\times 2$ table and additivity of $D_{ij}(u,v)$ imply that $\Delta_{ij}(u,v)=0$, for all $u,v$, and thus uniqueness follows.

In order to prove this property when one or both logit types are of type C or R, tables $\bd p(i,j)$ need to be considered in a specific sequence, since appropriate adjustments are required for the fixed marginal constraints to hold. Consider, for instance, the CC case, then the marginal constraints by row may be written as
$$
p_1(i;u) = \sum_{h=1}^{j-1} p(i,h;u,0)+p(i,j;u,+) =
\sum_{h=1}^{j-1} \tilde p(i,h;u,0)+\tilde p(i,j;u,+) = \tilde p_1(i;u);
$$
with the constraints by columns having a similar form. Thus, in order to implement these constraints, we must have already shown that $\sum_{h=1}^{j-1}[p(i,h;u,0)-\tilde p(i,h;u,0)]$ = 0. For this, we order cut points $i,j$ lexicographically and process the corresponding $2\times 2$ tables one at a time. For $i=j=1$, table $\bd p(1,1)$ is the same as in the GG case, so the marginal constraints hold without any adjustment. It can be easily seen that for any other $i,j$, once we have processed all tables preceding $\bd p(i,j)$ in the order mentioned above, we have already proved that
$$
\sum_{h=1}^{j-1}[p(i,h;u,0)-\tilde p(i,h;u,0)] = \sum_{h=1}^{i-1}[p(h,j;0,v)-\tilde p(h,j;0,v)] = 0,
$$
thus, the adjusted marginal constraints can be imposed.  The rest of the proof follows along the lines of the GG case. $\Box$
\subsubsection{Only one of the row and column logits different than L}
Assume, without loss of generality, that the row logits are of type L.
\begin{proposition}
Let $\bd p(j)$,  $j\in\{1,\ldots, I_2-1\}$, be the set of $I_1\times 2$ tables with entries as defined in (10)
and further consider logit type for rows L and for columns different than L. For fixed row marginals $\bd\pi_1$ and fixed column marginals
$\bd p_2(j)=(p(+,j;0), p(+,j;1))$, for given $j$, the vector $\bd\gamma(j)=(\gamma_{1j},\ldots, \gamma_{I_1-1,j})$ of $\phi$-scaled interactions $\gamma_{ij}$ of type Lh, $h \in \{G,C,R\}$, determines uniquely the tables $\bd p(j)$.
\end{proposition}
{\sc Proof}. Let us start from the LG case.
Consider, similarly to the propositions above, $\tilde{\bd p}(j) \neq \bd p(j)$ with $\bd\pi_1=\tilde{\bd\pi}_1$,
 $\bd p_2(j)=\tilde{\bd p}_2(j)$ and $\bd\gamma(j)=\tilde{\bd\gamma}(j)$ but with $\bd\upsilon(j) \neq \tilde{\bd\upsilon}(j)$, where  $\bd\upsilon(j)$ is an $I_1\times 2$ table with entries $\upsilon_{j}(i,v)=\sum_k \phi_k\mu_{k,i}\nu_{k,j+v}$.
 Note that the terms $\alpha_{ij}+\beta_{jv}$ in (10) and the differences $\upsilon_{j}(i,v)-\tilde\upsilon_{j}(i,v)$ have an additive structure with respect to $i,v$ for fixed $j$. In this context we may define
$$
\Delta_{j}(i,v)=\frac{p(i,j,v) - \tilde p(i,j,v)}{\pi_{i+}p(+,j,v)}  = G(\alpha_{ij}+\beta_{jv}+\upsilon_{j}(i,v))-
G(\alpha_{ij}^*+\beta_{jv}^*+\upsilon_{j}(i,v)).
$$
The marginal constraints are similar to those in the LL case except that the tables here have only two columns. Again, because the $G$ function is increasing, we can show the marginal constraints are not compatible with the additive structure of the argument of the same function, unless $\Delta_{j}(i,v)=0$ for all $i,j$.

The LC and LR cases need  minor adjustments in the marginal constraints by row
$$
\sum_{h=1}^{j-1}[p(i,h,0)-\tilde p(i,h,0)]+p(i,j,+)-\tilde p(i,j,+) = 0,
$$
analogue to the adjustments discussed in Proposition 2.
For them to apply, we need to start with $j=1$ and process sub-tables one at a time so that, when dealing with a given $j$, we have already shown that $\sum_{h=1}^{j-1}[p(i,h,0)-\tilde p(i,h,0)]=0$. $\Box$

The uniqueness of mappings shown above, implies that the extended RC(K) models (\ref{KaThe}), (\ref{GeKat}) and (\ref{GeKat2}), for appropriately fixed marginal totals, can equivalently be expressed in terms of the $\gamma_{ij}$s by (\ref{GeKat_gamma}).
\subsection{Positive association}
Results concerning positive dependence have been considered by \cite{Goodman81} for the ordinary RC(1) model, by \cite{gilula1988ordinal} for the RC(1) correlation model and by \cite{Bart02} for possible combinations of logit types, within the Kullback-Leibler divergence. In this section we investigate whether similar properties hold also when interactions are defined according to a general $\phi$ divergence measure. Such properties are relevant especially within an RC(1) model: the question is whether non decreasing row and columns scores, which imply $\gamma_{ij}(F;l_1,l_2)\geq 0$, imply also some kind of positive association between the row and column variables.

The common underlying theme in the results  that follow in this section
is an extension of the basic idea used  in the proof of \citet[Property 2]{rom1992generalized} to the much more general framework of the representation theorem. Before presenting the main results, it is convenient to introduce some notation and derive certain preliminary results. As a general rule, for pair of models obtained by swapping the logit type by row with the logit type by column, results will be derived for only one element of the pair, since those for the other follow by exchanging rows and columns. In addition, because models having one or both logit types equal to R may be converted into models with one or both logits of type C by reordering the categories of the corresponding variable, these models will not be examined explicitly as well.

For an $I_1\times I_2$ table of joint probabilities, $(\pi_{ij})$, define
$$
p_{ij}=\frac{\pi_{ij}}{\pi_{i+}},\:   s_{ij} =\frac{\sum_{h=j+1}^{I_2} \pi_{ih}}{\pi_{i+}},\:  \ \text{and} \:
S_{iju} = \frac{p(i,j;u,1)}{p_1(i,u)},
$$
where $p(i,j,u,v)$, $p_1(i,u)$ are the quadrant probabilities used to define the GG models, so, for instance, $p(i,j,1,1)$ = $\sum_{h=i+1}^{I_1}\sum_{l=j+1}^{I_2}\pi_{hl}$; both $s_{ij}$ and $S_{iju}$ are conditional survival function by row or after collapsing the original tables into a $2\times 2$ table, respectively.

\begin{lemma}
The log-odds ratios of type LG are all non negative if and only if $s_{ij}\leq s_{i+1,j}$, for all $i<I_1$, $j<I_2$. A sufficient condition for this to happen is that the difference $p_{ij}-p_{i+1,j}$, as $j$ goes from 1 to $I_2$, changes sign from positive to negative only once.
\label{LeLG}
\end{lemma}
{\sc Proof}. The first part of the statement corresponds to the well-known equivalence between $\cg P_D(L,G)$ and the simple stochastic ordering of the conditional distributions by row (see (3.2) of \cite{Grove1984}).
The second part of the statement holds trivially when $p_{ij}=p_{i+1,j}$ for all $j$; thus suppose that $p_{ij}-p_{i+1,j} \ne 0$ for some $j$, otherwise the result holds trivially.
Let $c$ be the smallest index such that $s_{ic}< s_{i+1,c}$ (i.e. $p_{ij}=p_{i+1,j}$ for $j<c$), then it can be easily verified that $p_{ic} - p_{i+1,c}>0$. Let $d$, $d>c$, be the smallest index such that $p_{id} - p_{i+1,d}<0$. Then, since $\sum_j (p_{ij}-p_{i+1,j})$ = 0, the sign of $s_{i+1,j}- s_{ij}$ for $j\geq d$ must remain non negative. $\Box$

Below we refer to the definition of $p(i,j;u,v)$  under the CC model and present two preliminary results.
\begin{lemma}
Suppose that $\gamma_{ij}$, the scaled interactions of type CC are all non negative, then
$$
s_{i,j} \leq s_{i+1,j},\: i<I_1, j<I_2.
$$
\label{sur}
\end{lemma}
{\sc Proof} Let
$$
d_{ijv} = \frac{p(i,j;0,v)}{p_1(i,0)}- \frac{p(i,j;1,v)}{p_1(i,1)}
$$
and note that $d_{ij1}$ = $s_{ij}-s_{i+1,j}$; because the function $G$ in the representation theorem is strictly increasing and the $\beta_{ijv}$ simplify, $d_{ijv}$ has the same sign as
$$
\Delta_{ijv} = \alpha_{ij0}-\alpha_{ij1}+\sum_k\psi_k (\mu_{k,i}-\mu_{k,i+1})\nu_{k,j+v};
$$
in addition, $\Delta_{ij1} - \Delta_{ij0}$ = $-\gamma_{ij}\leq 0$. To account for the fact that in the CC case for fixed $i,j$, the $2\times2$ table with entries $p(i,j;u,v)$, $u,v=0,1$, ignore columns to the left of $j$, it is convenient to start from $j=1$ and proceed, by induction, along the rows. For $j=1$, the conditional probabilities in the definition of $d_{ijv}$ sum to 1, thus $\sum_v d_{i1v}$ = 0, which implies that, unless $d_{i1v}$ = 0, $d_{i11} \leq 0$, so that $s_{i,1} \leq s_{i+1,1}$. For fixed $i$, using induction on $j$, it follows that $s_{i,j} \leq s_{i+1,j}$ for $j<I_2$. $\Box$
\begin{lemma}
Suppose that (i) $S_{i-1,j0}\leq S_{i-1,j1}$ and (ii) $s_{ij}\leq S_{ij1}$, then $S_{ij0}\leq S_{ij1}$.
\label{wav}
\end{lemma}
{\sc Proof}.
Let
$$
r_i=\frac{p_1(i,1)}{p_1(i-1,1)}, \quad t_i=\frac{p_1(i-1,0)}{p_1(i,0)};
$$
it can be easily verified that
$$
(A)\: S_{i-1,j1} = r_i S_{ij1}+(1-r_i) s_{ij} , \quad
(B)\: S_{ij0} = t_i S_{i-1,j0}+(1-t_i) s_{ij}.
$$
Replace $s_{ij}$ with $S_{ij1}$ in (A) above, then (ii) implies that
$S_{i-1,j1} \leq r_i S_{ij1} +(1-r_i) S_{ij1}$ = $S_{ij1}$. Next replace $S_{i-1,j0}$ with $S_{i-1,j1}$ and  $s_{ij}$ with $S_{ij1}$ in (B) above, then (i) and (ii) imply that
$S_{ij0} \leq t_i S_{ij1}+(1-t_i) S_{ij1}$ = $S_{ij1}$. $\Box$

\begin{proposition}
Suppose that at least one of the logit types is set to G and that all $F$ scaled interactions $\gamma{ij}\geq 0$, then the LOR of the same type are also non negative.
\label{ProPos}
\end{proposition}
{\sc Proof}. Let us start from the case of LG models, recall the expression (\ref{GeKat2}) for the $p(i,j,v)$ probabilities in Lemma 1 
and define
$$
d_{ijv} = \frac{p(i,j,v)}{\pi_{i+}}- \frac{p(i+1,j,v)}{\pi_{i+1,+}}.
$$
Because the $G$ function is increasing and the $\beta$s do not depend on $i$, it follows that $d_{ijv}$ has the same sign as
$$
\Delta_{ijv} = \alpha_{ij}-\alpha_{i+1,j}+\sum_k\psi_k (\mu_{k,i}-\mu_{k,i+1})\nu_{k,j+v};
$$
clearly, $\gamma_{ij}\geq 0$ implies that $\Delta_{ij1}\leq \Delta_{ij0}$. Because the two terms in the definition of $d_{ijv}$ are conditional probabilities, $\sum_v d_{ijv}$ = 0, thus, unless $d_{ijv}$ = 0, $d_{ij0}>0> d_{ij1}$, which implies $s_{ij}\leq s_{i+1,j}$. Then the result follows from Lemma \ref{LeLG}.

Now consider the case of GG models and, recalling the meaning of $p(i,j;u,v)$ in this case, define
$$
d_{ijv} = \frac{p(i,j;0,v)}{p_1(i,0)}- \frac{p(i,j;1,v)}{p_1(i,1)}.
$$
By using formula (\ref{GeKat}) from the representation theorem  and following the same arguments as above, it follows in this case that $d_{ijv}$ has the same sign as
$$
\Delta_{ijv} = \alpha_{ij0}-\alpha_{ij1}+\sum_k\psi_k (\mu_{k,i}-\mu_{k,i+1})\nu_{k,j+v}.
$$
Because the rations appearing in the definition of $d_{ijv}$, $p(i,j;u,v)/p_1(i,u)$, are conditional  probabilities summing to 1 with respect to $v$,  $d_{ij0}+ d_{ij1}$ = 0,
thus, either the two addends are both 0, or $d_{ij1}<0$, which implies $S_{ij0}\leq S_{ij1}$, a necessary and sufficient condition for quadrant dependence.

For the case of CG models, define $d_{ijv}$ as in the GG case above, but note that now
$\frac{p(i,j;0,v)}{p_1(i,0)}$, $v=0,1$ are conditional probabilities along the $i$th row. By similar arguments as before, it can be verified that $d_{ijv}$ has the same sign as
$$
\Delta_{ijv} = \alpha_{ij0}-\alpha_{ij1}+\sum_k\psi_k (\mu_{k,i}-\mu_{k,i+1})\nu_{k,j+v}
$$
and furthermore that
$$
\frac{p(i,j;0,1)}{p_1(i,0)}\leq \frac{p(i,j;1,1)}{p_1(i,1)},
$$
which is a necessary and sufficient condition for the LOR of type GC to be non negative. $\Box$

When neither the row nor the column logit are of type global there exist tables of probabilities such that, while $\gamma_{ij}(F,l_1,l_2)\geq 0$ for all $i<I_1,\: j<I_2$, at least for a pair of $i,j$, $\eta_{1,2,l_1,l_2}(i,j)<0$.
Examples of this kind will be provided for the Cressie-Read $\lambda$-scaled interactions $\gamma_{ij}(\lambda,l_1,l_2)$ and the main pairs of logit types (LL, LC and CC) in the Appendix.
Because setting the logit type by row to R is the same as assigning logit type C and reversing the rder of the row categories, examples involving logit type R may be derived from those with logit type C. Some examples in the opposite direction, showing that
$\eta_{1,2,l_1,l_2}(i,j)>0$ does not imply $\gamma_{ij}(\lambda,l_1,l_2)>0$ were presented in \cite{EspeM}, p. 143.

A direct consequence of Proposition \ref{ProPos} is that $F$-scaled RC(K) models for several logit pairs (GG, GL, LG, GC, CG, GR, RG) with all scaled interactions  $\gamma_{ij}(F,l_1,l_2)\geq 0$,
imply that the corresponding log-odds ratios are also non negative and thus stochastic orderings of the corresponding kind also holds.

The results of Proposition \ref{ProPos} do not apply to nine types of $F$ scaled obtained by combininng logit types L, C, R in all possible pairs. For these types of logits, weaker notions of positive dependence hold, as shown below.
The next result is a slight extension of \citet[Proposition 4]{Kateri95} from RC(1) to RC(K) models. The proof is essentially based on the same method used by \cite{rom1992generalized} in their Proposition 2 which refers to RC(1) models and a subclass of the scaled interactions considered here.
\begin{proposition}\label{LL_stoch_ord}
Suppose that the RC(K) model (\ref{KaThe}) holds and that the $F$-scaled interactions of type LL
$\gamma_{ij}$ are all non negative, then the log-odds ratios of type L-G and G-L are also non negative.
\end{proposition}
{\sc Proof}.
Let us consider the LG case first and suppose that the difference $p_{ij}-p_{i+1,j}$ is not identically 0, otherwise the result is obvious. Note that
$$
p_{ij}-p_{i+1,j} = \pi_{+j}[ G(\alpha_i+\beta_j+\sum_k\psi_k\mu_{ki} \nu_{kj}) -
 G(\alpha_{i+1}+\beta_j+\sum_k\psi_k\mu_{k,i+1} \nu_{kj}).
$$
Due to the fact that function $G$ is strictly increasing and $\pi_{+j}$ is positive, the sign of $p_{ij}-p_{i+1,j}$ is the same as the sign of
$$
\Delta_{ij}=\alpha_i-\alpha_{i+1}+\sum_k\psi_k(\mu_{ki}-\mu_{k,i+1})\nu_{kj}.
$$
However, under the given setting, $\Delta_{i,j+1}-\Delta_{ij}$ = $-\gamma_{ij}\leq 0$. Hence, for fixed $i$, the $\Delta_{ij}$ are decreasing in $j$, thus, if they are not identically 0, they can change sign only once. Since this must hold also for $p_{ij}-p_{i+1,j}$, the result follows from Lemma \ref{LeLG}. The proof that the log-odds rations of type G-L are also non negative follows by exchanging rows with columns and using the same arguments as above. $\Box$

It is worth noting that while the assumption of non negative LOR of type LL implies the strongest notion of positive dependence, the condition of non negative scaled interactions of type LL is weaker as it implies non negative LOR of type LG, GL and, as a consequence, GG.
\subsubsection{From LC to LG}
\begin{proposition}
Non negative $F$-scaled interactions of type LC imply non negative LORs of type LG and GG.
\end{proposition}

{\sc Proof}.
Recall the form of the representation theorem in the LC case
$$
p(i,j,v) = \pi_{i+} p(+,j,v) G\left(\alpha_{ij}+\beta_{jv}+
\sum_k\psi_k\mu_{k,i}\nu_{k,j+v}\right);
$$
for $i< I_1$, $j<I_2$ and $v\in (0,1)$, and let
$$
d_{ijv} = \frac{p(i,j,v)}{\pi_{i+}}-\frac{p(i+1,j,v)}{\pi_{i+1,+}}
$$
be the difference between the conditional distributions in two adjacent rows.
By an argument similar to the one in Proposition \ref{LL_stoch_ord}, because $G$ is strictly increasing and the $\beta_{jv}$s simplify, the sign of $d_{ijv}$ is the same as the sign of
$$
\Delta_{ijv} = \alpha_{ij}-\alpha_{i+1,j}+\sum_k\psi_k(\mu_{k,i}-\mu_{k,i+1})\nu_{k,j+v}.
$$
Because $\Delta_{ij1} - \Delta_{ij0}$ = $-\gamma_{ij}$,  $\gamma_{ij}\geq 0$  implies $\Delta_{ij1} \leq \Delta_{ij0}$. Recall that $S_{ij}$ = $p(i,j,1)/\pi_{i+}$ is the survival function for the conditional distribution in row $i$.
Starting from $j=1$, because $p(i,1,0)/\pi_{i+}+S_{i1}$ = 1 for $i\in \cg I$, if the $d_{i1v}$ are different from 0, they must be of different signs and, due to $\Delta_{i10}\geq\Delta_{i11}$, $d_{i11}$ must be negative, implying that $S_{i1}\leq S_{i+1,1}$. Going by inductions, and assuming that the inequality holds for $j-1$, i.e. $S_{i,j-1}\leq S_{i+1,j-1}$, to prove that $S_{i,j}\leq S_{i+1,j}$, for $j<I_2$, let us suppose that the inequality is reversed. Then $d_{ij1}$ would have to be positive which implies also $d_{ij0}>0$, which is impossible because it would mean that the two addends in the left hand side of the inequality below are greater than the corresponding addends on the right hand side
$$
\frac{p(i,j,0)}{\pi_{i+}}+S_{ij}=S_{i,j-1}\leq S_{i+1,j-1} = \frac{p(i+1,j,0)}{\pi_{i+1,+}}+
S_{i+1,j}.
$$
Because non negative LOR of type LG imply non negative LOR of type GG, the second part of the statement follows. $\Box$
\subsubsection{From CC to GG}
Below refer to the definition of $p(i,j;u,v)$ in the CC model.
\begin{proposition}
If the scaled interacitons of type CC are all non negative,  the LOR of type GG are all non negative.
\end{proposition}
{\sc Proof}. Lemma \ref{sur} implies that the assumption of Lemma \ref{wav} holds, that is, each $2\times 2$ table resulting from collapsing the original table at $i<I_1,j<I_2$, has the conditional  survival function larger in the second row. $\Box$
\section{Maximum likelihood estimation}
Given a table of observed frequencies $\bd y$, a model $\cg M$ is defined by choosing the logits type
for the row and column variables, the value of the power divergence $\lambda$, the rank $K$ of the RC
model and, possibly, additional linear constraints on the marginal logits and the interactions. Maximum
likelihood may be computed by the \cite{aitchison1958maximum} algorithm where, in each step, a
quadratic approximation of the log-likelihood is maximized subject to a linear approximation of the
constraints. Below, a special version of this algorithm suitable in the present context is recalled
briefly.
\subsection{The regression algorithm}
Let $\bd\theta$ denote a vector of canonical parameters for the multinomial distribution; this may be
defined by a full rank matrix $\bd G$ of size $I_1 I_2 \times(I_1 I_2-1)$ whose columns do not span the
unitary vector, then $\log \bd\pi$ = $\bd G\bd\theta-\bd 1\log(\sum\exp(\bd G\bd\theta))$. Suppose that
the model $\cg M$ is defined by the vector of constraints
$\bd h(\bd\theta)=\bd 0$; let $\bd H$ be the derivative of $\bd h\tr$ with respect to $\bd\theta$; let
also $\bd H^-$ = $\bd H(\bd H\tr\bd H)^{-1}$ denote a right inverse of $\bd H\tr$. A first order
approximation of the constraints may be written as
\begin{equation}
\bd h=\bd h_0+\bd H\tr(\bd\theta-\bd\theta_0)=\bd H\tr(\bd H^-\bd h_0+\bd\theta-\bd\theta_0) = \bd
H\tr\bd v=\bd 0.
\label{lapp}
\end{equation}
Let $\bd X$ be a matrix that spans the space orthogonal to the columns of $\bd H$, then (\ref{lapp})
implies $\bd v$ = $\bd X\bd\beta$, say. Let $Q$ denote the quadratic approximation of the
log-likelihood at $\bd\theta_0$ which has the sam score and the same expected information, then the
least square solution that maximizes $Q$ subject to the linear approximation of the constraints in
(\ref{lapp}) has the form
$$
\hat{\bd v} = \bd X(\bd X\tr\bd F_0\bd X)^{-1}\bd X\tr\bd F_0(\bd H^-\bd h_0+\bd F_0^{-1}\bd s_0),
$$
where $\bd s$ = $\bd G\tr(\bd y-n\bd\pi)$ and $\bd F$ = $n\bd G\tr(\diag(\bd\pi)-\bd\pi\bd\pi\tr)\bd G$
denote, respectively, the score and information matrix.
From this, an updated estimate along the line determined by the least square solution may be computed
as $\bd\theta(t)$ = $\bd\theta_0+t(\hat{\bd v}-\bd H_0^-\bd h_0)$, where $t$ denotes the step length.
Clearly the value of $t$ should be such that the likelihood increases and $\bd h(\bd\theta(t))$ is
possibly smaller that $\bd h(\bd\theta_0)$.
\subsection{Implementing the rank constraint}
To use the regression algorithm, we need to translate the constraint that the matrix $\bd M$ of $F$-scaled interactions has rank $K$  into the constraint that a set of non linear functions of the interactions are equal to 0. One may use the following result proven by \cite{Bart02}:
suppose that, for some $i,j$, $m_{ij}\ne 0$, let $\bd m_{1i}$ be the column vector
containing  the $i$th row of $\bd M$ and $\bd m_{2j}$ its $j$th column; let also $\bd H_{\ell v}$ be the
identity matrix of size $I_{\ell}-1$ without the $v$th row, $\ell=1,2$.
\begin{lemma}
Let $\bd F(\bd M)$ = $\bd H_{1i}(\bd M-\bd m_{1i}\bd m_{2j}\tr/m_{ij})\bd H_{2j}\tr$; if $\bd M$ has
rank $K$, then $\bd F(\bd M)$ has rank $K-1$.
\label{rank}
\end{lemma}
{\sc Proof}
See \cite{Bart02}, Appendix.

Because the transformation in Lemma \ref{rank} may be applied recursively,
let $\bd F^r(\bd M)$ denote $r$ applications of the same transformation; if
rank$(\bd M)$ = $K$, then Lemma \ref{rank} implies that $\bd F^K(\bd M)$ is a
matrix of zeros. It follows that, under the assumed model, the vectorised
version of the matrix of interactions may be transformed into a vector whose
elements, under model (\ref{Goodman}) is a vector of 0s.
This approach has two convenient features:
\begin{enumerate}
\item the result in Lemma \ref{rank} is independent of the type of interactions adopted and thus it
    can be applied to the extended interactions defined above;
\item the derivative of $\bd F^r(\bd M)$ with respect to $\bd M$, in its vectorised form, may be
    computed by a simple recursive formula.
\end{enumerate}
\subsection{Line search}
Though one could choose the step length $t$ by trial, the approach described here can improve efficiency.
It consists in computing the value of the function
$$
f(t) = \bd y\tr\log(\bd\pi(t))/n-\bd h(t)\tr\bd h(t)/2
$$
computed at $t$ = 0, 1/4, 1/2 and maximizing the cubic approximation of $f(t)$ which uses also its
first derivative at $t=0$.

The algorithm described above can be implemented in the {\ttfamily R}-package {\ttfamily extRC}
of \cite{Bartolucci2020}, which is also used for the analysis of the application that follows.
\section{Application}
The data in the left side of Table \ref{tab:1} is taken from \cite{Mosteller1968} and classifies 3,500
British adults according to their social class and that of their fathers.

\begin{figure}[hb]
\centering
 \includegraphics[width=0.75\textwidth]{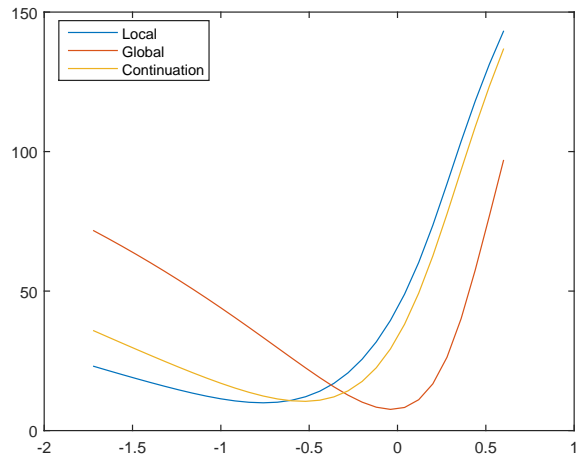}
\caption{\label{Fig1} \it Deviance for a collection of RC models as a function of $\lambda$ for
different choices of logit type.}
\end{figure}

\begin{table}[hb]
\caption{\label{tab:1} Left panel: occupational status (OS) of father ($F_i$) and son $(S_j)$ for 3,500
British adults; Right panel: estimated cumulative conditional distributions of son's OS conditional on
that of the father}
\vskip1mm
\centering
\fbox{
\begin{tabular}{lcrrrrrcrrrr}
 \hline
   & \hspace{1mm} & $S_1$ & $S_2$ & $S_3$ & $S_4$ & $S_5$ &\hspace{1mm} & $S_1$ & $S_2$ & $S_3$ & $S_4$
   \\
 \hline
$F_1$ &&  50 &  45 &   8 &  18 &   8  && 0.3932 & 0.7188 & 0.7859 & 0.9457 \\
$F_1$ &&  28 & 174 &  84 & 154 &  55  && 0.0560 & 0.3852 & 0.5670 & 0.8906 \\
$F_1$ &&  11 &  78 & 110 & 223 &  96  && 0.0159 & 0.1747 & 0.3683 & 0.8214 \\
$F_1$ &&  14 & 150 & 185 & 714 &  447 && 0.0106 & 0.0970 & 0.2311 & 0.7128 \\
$F_1$ &&   3 &  42 &  72 & 320 &  411 && 0.0058 & 0.0516 & 0.1314 & 0.5298 \\
\hline
\end{tabular}}
\end{table}
As a preliminary step, a quick search among possible model specifications with $K=1$ was performed;
because the row and column variables have the same nature, it seems reasonable to restrict attention to
models with the same logit type for $X_i,\:X_2$. The deviance of models with logit type {\it local},
{\it global} and {\it continuation} as a function of $\lambda$ are plotted in Figure \ref{Fig1}. The
model with logits of type global with $\lambda$ = -0.04 seems to do best; however, for a range of
$\lambda$s close to 0, difference in model fit are not substantial.
Both the models with equally spaced row and column scores are rejected as well as the model of
marginbal homogeneity. However, the model which assumes a constant shift between the corresponding
marginal logits is not rejected ($P$ value 14.30\%); it implies that the distribution of occupational
status of the sons is stochastically larger relative to that of the fathers. Taking this as the final
model, $\hat\psi$ = 1.98 and the correlation between $X,Y$ based on the estimated scores equals 0.46.
This seems to be in accordance with the fact that, being the estimated row and column scores
increasing, all elements of the estimated matrix of extended interactions are positive. The estimated
values of the cumulative conditional distributions given in Table \ref{tab:1} indicate clearly that,
the estimated model, satisfies positive quadrant dependence.

\begin{table}[h]
\caption{\label{tab:2} Left side: deviance and degrees of freedom (dof) for a selection of models with
$K=1$, $\lambda=-0.04$ and logit type global for row and column variable; Right panel: estimated row
and column scores for the RC model with shift on marginal logits}
\vskip1mm
\centering
\fbox{
\begin{tabular}{lrrcrr} \hline
   & Deviance & dof & \hspace{1mm} & row scores & column scores\\
 \hline
RC & 7.60 & 9       &&  -2.8343 & -2.9825\\
R & 50.15 & 12      &&  -1.5076 & -1.6513\\
C & 55.88 & 12      &&  -0.5235 & -0.6150\\
RC+m.h. & 40.47 & 13&&   0.3199 &  0.2474\\
RC+m.s & 17.19 & 12 &&   1.0738 &  1.0162
\\
\hline
\end{tabular}}
\\
{\scriptsize m.h. = marginal homogeneity, m.s. = constant shift on marginal logits}
\end{table}
\section*{Appendix}
The details of the three counter examples showing that positive  Cressie-Read $\lambda$-scaled interactions of type LL, LC and CC do not imply positive LOR for the same pair of ligit types are given in the tables below.
\begin{table}[h]
\caption{\label{tab:3} A $3\times 3$ table of probabilities (left side) where $\gamma_{ij}(\lambda ,L,L)$, $\lambda=7$, are all non negative but one $\eta_{1,2,L,L}(i,j)$ is negative}
\vskip1mm
\centering

\begin{tabular}{rrrcrrcrr} \hline
\multicolumn{3}{c}{$\pi_{ij}$} && \multicolumn{2}{c}{$\gamma_{ij}(\lambda,L,L)$} && \multicolumn{2}{c}{$\eta_{1,2,L,L}(i,j)$} \\
\hline
0.1444 & 0.1018 & 0.0939 & & 0.8100 & 0.0900 & & 0.3365 & 0.8109\\
0.0979 & 0.1117 & 0.1175 & & 0.0810 & 0.0090 & & 0.2136 & -0.0225\\
0.0914 & 0.1178 & 0.1236 & &       &         & &       &
\\
\end{tabular}
\end{table}

\begin{table}[h]
\caption{\label{tab:4} A $3\times 3$ table of probabilities (left side) where $\gamma_{ij}(\lambda ,L,C)$, $\lambda=5$,  are all non negative but one $\eta_{1,2,L,C}(i,j)$ is negative}
\vskip1mm
\centering
\begin{tabular}{rrrcrrcrr} \hline
\multicolumn{3}{c}{$\pi_{ij}$} && \multicolumn{2}{c}{$\gamma_{ij}(\lambda ,L,C)$} && \multicolumn{2}{c}{$\eta_{1,2,L,C}(i,j)$} \\
\hline
0.1418 & 0.1064 & 0.0355 & & 0.3839 & 0.4860 & & 0.3365  & 0.8109\\
0.1773 & 0.1418 & 0.1064 & & 0.1980 & 0.0740 & & 0.2136  & -0.0225\\
0.1064 & 0.1064 & 0.0780 & &       &         & &         &
\\
\hline
\end{tabular}
\end{table}

\begin{table}[h]
\caption{\label{tab:5} A $3\times 3$ table of probabilities (left side) where $\gamma_{ij}(\lambda ,C,C)$,  $\lambda=16$), are all non negative but one $\eta_{1,2,C,C}(i,j)$ is negative}
\vskip1mm
\centering
\begin{tabular}{rrrcrrcrr} \hline
\multicolumn{3}{c}{$\pi_{ij}$} && \multicolumn{2}{c}{$\gamma_{ij}(\lambda ,C,C)$} && \multicolumn{2}{c}{$\eta_{1,2,C,C}(i,j)$} \\
\hline
 0.1695 & 0.0847 & 0.0847 & & 0.0518 & 0.2042 & & 0.0513 & 0.2007\\
 0.1525 & 0.0678 & 0.0847 & & 0.0973 & 0.0082 & & 0.0953 & -0.0408\\
 0.1695 & 0.0847 & 0.1017 & &        &        & &        &
\\
\hline
\end{tabular}
\end{table}

\subsection*{Acknowledgments}
The first author would like to thank Francesco Bartolucci and Yosi Rinott for discussions.

\bibliography{RC2}

\end{document}